# Modelling butterfly wing eyespot patterns

**Rui Dilão**[1,2] and **Joaquim Sainhas**[1,3]

1) Grupo de Dinâmica Não-Linear, Instituto Superior Técnico, Departamento de Física, Avenida Rovisco Pais, 1049-001 Lisboa, Portugal.

2) Institut des Hautes Études Scientifiques, Le Bois-Marie, 35, route de Chartres, F-91440 Bures-sur-Yvette, France.

3) Faculdade de Motricidade Humana, Estrada da Costa, 1495-688 Cruz Quebrada, Lisboa, Portugal.

**Abstract**

Eyespots are concentric motifs with contrasting colours on butterfly wings. Eyespots have intra- and inter-specific visual signalling functions with adaptive and selective roles. We propose a reaction-diffusion model that accounts for eyespot development. The model considers two diffusive morphogens and three non-diffusive pigment precursors. The first morphogen is produced in the focus and determines the differentiation of the first eyespot ring. A second morphogen is then produced, modifying the chromatic properties of the wing background pigment precursor, inducing the differentiation of a second ring. The model simulates the general structural organisation of eyespots, their phenotypic plasticity and seasonal variability, and predicts effects from microsurgical manipulations on pupal wings as reported in the literature.

The most remarkable phenotypic features of butterflies are their wing patterns, which have adaptive functions ranging from passive camouflage and mimicry to competitive self-advertising (Nijhout 1991). They display great plasticity and adaptability to seasonal and environmental changes (Smith 1978; Koch 1992). Butterfly wing eyespots, or *ocelli*, are concentric rings with intense and contrasting coloration. They mimic the global appearance of vertebrate eyes, and have active signalling functions against predators, as well as sexual signalling roles for mate competition (Nijhout 1981; Nijhout 1996; Vane-Wright & Boppré 1993). Selection studies have shown eyespot patterns reveal high heritability (Brakefield & Noordwijk 1985; Brakefield & French 1999; Monteiro *et al.* 1997). The biology of butterfly wing patterns involves different levels of description, ranging from molecular mechanisms of development to selection and pattern evolution. In fact, the link between developmental patterns and evolution processes is an important biological problem (Nijhout 1991; Brakefield & French 1999).

Insect wings have two independent global morphogenic fields, corresponding to the epithelial monolayers of the wing surfaces. As lepidopteran patterns are different in both wing surfaces, morphogenic signals responsible for pattern formation propagate within each monolayer through gap-junctions, facilitating diffusive intercellular communication (Unwin & Zampighi 1980; Nijhout 1990). Global wing patterns result from the combination of a small number of elementary motifs. On a local scale, the areas delimited by wing veins correspond to independent morphogenic fields, where elementary patterns such as ocelli emerge. Wing patterns have a discrete structure formed by overlapping scales distributed along parallel rows. The scales are flattened protrusions of specialised epithelial cells that differentiate among homogeneous pigmentary colours or light diffracting microstructures (Nijhout 1991; Vukusic *et al.* 1999).

Some aspects of the genetic mechanisms of eyespot ontogeny have been clarified (Brakefield & French 1999; Beldade & Brakefield 2002). Eyespot development begins at the larval stage, and is initiated by the expression of wing-patterning genes at predetermined points (foci) of the developing wings (Beldade & Brakefield 2002; Carroll 1994; Brakefield *et al.* 1996). These foci create positional information that regulates the developmental pathway of eyespots, defining a cellular territory that corresponds to the emergent eyespot (Nijhout 1996; Nijhout 1994; French & Brakefield 1995; Brakefield & French 1995; French & Brakefield 1992).

In experiments with *Bicyclus anynana*, the inducing action of focus cells in eyespot formation is corroborated by surgical experiments that involve the grafting and destruction of focus cells in early pupal stages. In the first case, eyespot development occurs in the new unprogrammed areas. In the second case, an inhibition of normal eyespot development is observed (Nijhout 1994; French & Brakefield 1995). Both situations show the role of the eyespot focus as a source of a diffusing morphogenic signal that triggers the developmental pathway of eyespots (French & Brakefield 1995; Brakefield & French 1995; French & Brakefield 1992; Brunetti *et al.* 2001).

We propose the action of a minimum of two diffusive signalling substances, or morphogens (Turing 1952) as a model for eyespot development. The first morphogen $M_1$ is produced in a specific region of the wing by an initial precursor $A$. The spatial distribution of the initial precursor $A = A(x, y, t = 0)$ at time zero defines the localisation of foci. The differentiation of the first ring (dark region) and the production of the second morphogen $M_2$ are induced by $M_1$ reacting with a wing background pigment precursor $P_0$ and modifying its chromatic properties. The second morphogen $M_2$ reacts with the wing background pigment precursor $P_0$, generating a second ring or light aureole. The final pigmentation is always defined by the highest local concentration of the pigment precursors.

This model can be implemented with the following kinetic pathways: 1) Focal cells release a primary diffusive morphogen $M_1$. The simplest conceivable kinetic mechanism is the following: $A \xrightarrow{k_1} M_1$, $M_1 \xrightarrow{k_2}$, where $k_1$ and $k_2$ are reaction rates, $A$ is the morphogen precursor, and where the second reaction represents morphogen degradation. 2) The morphogen $M_1$ reacts with the background pigment precursor $P_0$ in the surrounding wing area, producing a new pigment precursor $P_1$ and a secondary morphogen $M_2$: $M_1 + P_0 \xrightarrow{k_3} P_1 + M_2$, $M_2 \xrightarrow{k_4}$. The pigment precursor $P_0$ is responsible for wing background pigmentation and $P_1$ for pigmentation of the first ring. 3) The diffusive morphogen $M_2$ produces a chemical modification in the pigmentation of the wing background: $M_2 + P_0 \xrightarrow{k_5} P_2$, where $P_2$ is the pigment precursor of the second ring. In figure 1, we schematically show this developmental mechanism.

At the beginning of the developmental process, the primary pigment determinant $P_0$ is spatially distributed along the wing surfaces, except at focal cells. Distribution of the foci is described by the spatial distribution of the morphogen precursor $A$: focal areas are characterised by positive values of $A$, whereas outside focal areas $A = 0$. All other pigment determinants $P_1$ and $P_2$ are absent at the beginning of the developmental process.

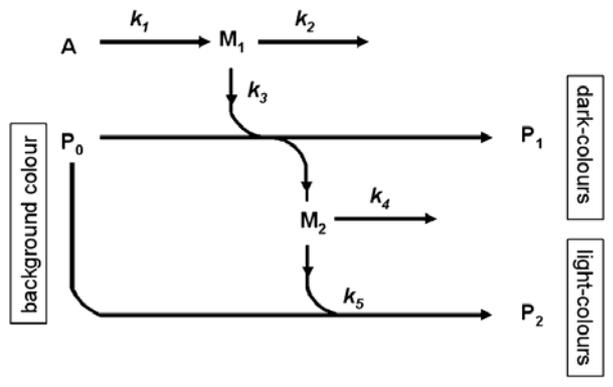

**Figure 1**. Schematic diagram showing the several kinetic mechanisms considered in our model for the formation of butterfly wing eyespot patterns. Morphogens or diffusive substances are represented by $M_1$ and $M_2$; $P_0$, $P_1$, and $P_2$ are pigment precursors; $k_1$ to $k_5$ are the rate constants for each elementary process. Applying the mass action law to the elementary processes and adding diffusion to signalling substances or morphogens, we obtain equations (1), that describe the space and time evolution of eyespot development. In this mechanism, pigment precursors may be considered as belonging to the same chemical category.

Applying mass action law formalism to the above elementary reactions, we obtain the following system of partial differential equations:

$$\frac{\partial A}{\partial t} = -k_1 A$$

$$\frac{\partial M_1}{\partial t} = k_1 A - k_2 M_1 - k_3 M_1 P_0 + D_1 \nabla^2 M_1$$

$$\frac{\partial M_2}{\partial t} = k_3 M_1 P_0 - k_4 M_2 - k_5 M_2 P_0 + D_2 \nabla^2 M_2$$

$$\frac{\partial P_0}{\partial t} = -k_3 M_1 P_0 - k_5 M_2 P_0 \quad (1)$$

$$\frac{\partial P_1}{\partial t} = k_3 M_1 P_0$$

$$\frac{\partial P_2}{\partial t} = k_5 M_2 P_0$$

where $D_1$ and $D_2$ are the diffusion coefficients of morphogens $M_1$ and $M_2$ on the two-dimensional wing region, $k_1$ to $k_5$ are the rate constants of the elementary kinetic processes, and $\nabla^2 = (\partial^2/\partial x^2 + \partial^2/\partial y^2)$ is the two-dimensional Laplace operator. We consider that the pigment precursors $P_i$ do not diffuse and that local wing colour is determined by the pigment precursor with the highest concentration.

It follows from system (1) that pigment precursors obey a mass conservation law of the form $P_0(x,y,t) + P_1(x,y,t) + P_2(x,y,t) = \text{constant}$, and that the spatial dependence of the pigment precursors is induced by diffusion of the morphogens. In the steady state ($t \to \infty$), the concentrations of morphogens $M_1$ and $M_2$ vanish, and stable patterns emerge, obeying the conservation law $P_0(x,y,t=0) = P_0(x,y,t=\infty) + P_1(x,y,t=\infty) + P_2(x,y,t=\infty)$. The spatial distribution of the pigment precursors may be considered a pre-pattern that determines subsequent patterning events. The model presumes a competitive selection of exclusive pigment synthesis pathways within scale cells, which agrees with the experimental results (Koch 1991; Koch & Kaufmann 1995).

To test this model, we numerically integrate equations (1) with an explicit finite-difference method, minimising the global numerical error and avoiding spurious symmetries induced by the two-dimensional integration lattice (Dilão & Sainhas 1998). The simulations were performed with zero flux boundary conditions in a two-dimensional spatial region of *101×101* cells. The simulations start with *A>0* and *P₀=0* at focal points. We set the uniform concentration of the primary pigment precursor away from focal points to be $P_0 > 0$, and $A = 0$. At time zero, all other variables in the model are null over the entire spatial range. If we let $P_0 > 0$ at focal points, the patterns emerging from the steady state solutions of equations (1) remain unchanged, but focal points are not differentiated. In fact, in different butterfly species, focal cells of ocelli can be white or unspecified. In figure 2, we depict the evolution of morphogens and pigment precursors according to the model equations (1).

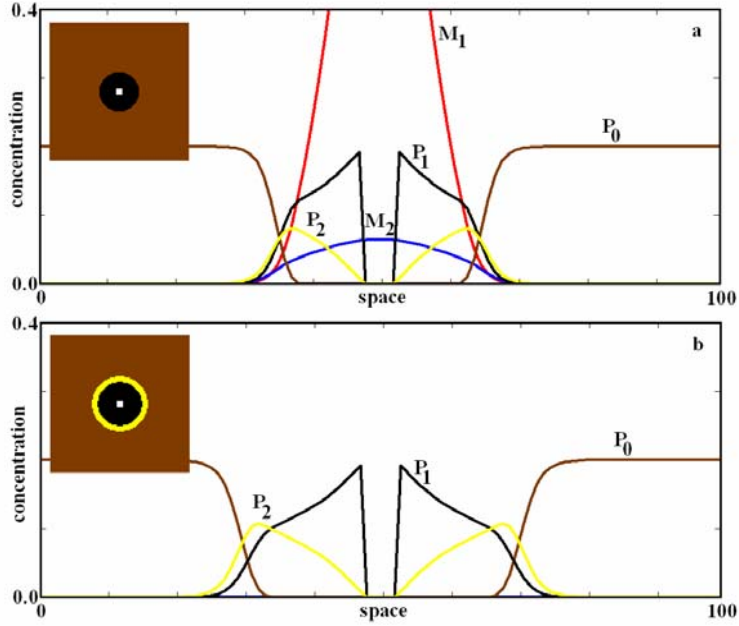

**Figure 2**. Solution of the reaction-diffusion system of equations (1) in a square of $101 \times 101$ cells for time $t = 15$ (**a**), and the steady state ($t \rightarrow \infty$) (**b**). The graphs represent the concentration profiles of the model variables taken along a cross-section passing through the focus. All variables are represented in dimensionless form. The squares on the top left indicate the concentration levels of pigment precursors. The highest $P_0$ concentration areas are represented in brown. Black and yellow represent the highest concentrations of $P_1$ and $P_2$, respectively. The white spots represent focal points. Patterns emerge as the steady state solutions of equations (1) for the pigment precursors. At the steady state, the morphogen concentrations are zero. Simulations were carried out with zero flux boundary conditions and the following initial conditions: $P_1(x, y, 0) = P_2(x, y, 0) = 0$ and $M_1(x, y, 0) = M_2(x, y, 0) = 0$. The focus region corresponds to a square of $5 \times 5$ cells and is characterised by the initial conditions $A(x, y, 0) = 20$ and $P_0(x, y, 0) = 0$. Outside the focal region, $A(x, y, 0) = 0$ and $P_0(x, y, 0) = 0.2$. We selected the following parameters: $k_1 = 1.0$, $k_2 = 0.05$, $k_3 = 4.0$, $k_4 = 0.01$, $k_5 = 4.0$, $D_1 = D_2 = 1.0$, $\Delta t = 0.1$ and $\Delta x^2 = 6 D_1 \Delta t$.

During the growth of the first ring (dark region), the morphogen concentration $M_2$ increases, promoting the formation of the second ring, figure 2. The grafting and destruction of focal cells are modelled by changing the initial concentration of the morphogen precursor $A$. In case of the destruction of focal cells, $A = 0$, wing colour is determined by the wing background pigment precursor, $P_0(x,y,t=0) = P_0(x,y,t=\infty)$, and no eyespots appear. Grafting of focal cells is modelled by the local introduction of $A$. The light ring increases with the area of the focal region, as observed in the wing patterns of different butterfly species, figures 3a-3c. In this model, it is the initial distribution of the morphogen precursor $A$ that determines phenotypic differentiation of butterfly wings, one of the main features associated with natural selection.

Eyespot variability may be understood within this framework. For example, by changing the position and width of a focus, it is possible to simulate different structures for eyespot patterns, as in figures 3d and 3e. Seasonal polyphenism and phenotypic plasticity are simulated by changing the rate constants of the model equations. In nature, this adaptability depends on environmental factors, such as temperature, relative humidity, and photoperiod (Nijhout 1991; Smith 1978; Koch 1992), changing the reaction rates of kinetic mechanisms. Generalising the cascading model presented here, additional concentric rings may be modelled by the inclusion of new morphogens and pigment precursors.

One of the novelties of this model is based on the assumption that the reaction-diffusion mechanism associated with the morphogens is the process that triggers the synthesis of the non-diffusive pigment precursor responsible for patterning. In the steady state, the concentration of morphogens is zero. This contrasts with the classical Turing reaction-diffusion approach, where patterning is due to spatial distribution of the morphogens at the steady state (Nijhout 1991; Turing 1952; Murray 1989).

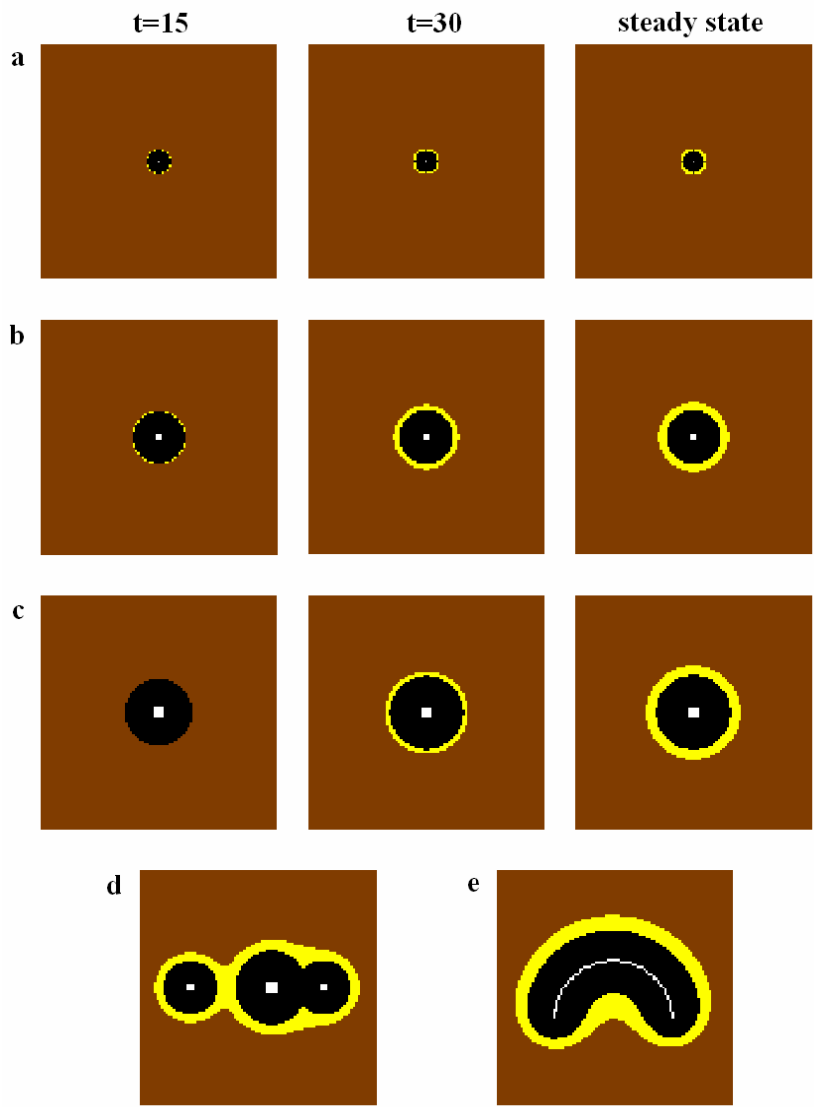

**Figure 3**. From **a** to **c** we depict the evolution of eyespot patterns over time as a function of the focal area for the same parameters values as in figure 2. In **a**, the focus region corresponds to one cell; in **b** to a square of $3\times 3$ cells, and in **c** to a square of $5\times 5$ cells. According to experimental observations, the larger the focus area, the larger the eyespot. In **d** and **e** we show that the variability and typical interaction patterns of eyespots are those predicted by the model equations (1) for point (**d**) and arc (**e**) foci. The patterns in **d** and **e** are steady state solutions of equations (1).

The classical model of pattern formation in butterflies is based on the assumption of simple diffusion and threshold mechanisms of pigments (Nijhout 1990; Murray 1989; Monteiro *et al.* 2001). According to Nijhout (2001, pp. 221), "[p]igment bands are formed at several threshold levels of a symmetrical gradient of pigment determination", and "each threshold becomes the origin of a new symmetrical gradient of pigment determination". In the model presented here, pigments are non-diffusive substances, morphogens are diffusive and reacting substances (Turing 1952), and we have no threshold assumptions. On the other hand, one of the undetermined issues of the classical model (Monteiro *et al.* 2001) is whether the focal signal is a constant level or constant rate source of morphogen. In the model presented here, the focal signal is determined by a localised morphogen precursor ( $A$ ) that decays during the developmental process.

In conclusion, we have shown that the development of butterfly wing patterns may be explained by a reaction-diffusion mechanism with two diffusive morphogens, triggering different pathways from a common pigment precursor $P_0$, in a cascading process. At the beginning of the developmental process, the pigment precursor $P_0$ is equally distributed along pupal wings, and we have considered that its dynamics is non-diffusive. This is in agreement with the chemistry of the major pigments in butterflies (Nijhout 1991). Changing the kinetic parameters of the model and the initial distribution of the morphogen precursor $A$, results in an abundant diversity of patterns, as observed in lepidopteran wings. This model consistently predicts the major features of the development of butterfly eyespot patterns, their structural organisation, phenotypic plasticity, seasonal variability, and the effects of microsurgical manipulations.

**Acknowledgements**

We thank Noah Hardy for carefully reading and editing the text. This work has been partially supported by *Fundação para a Ciência e a Tecnologia* (Portugal), under the framework of the POCTI Project /FIS/13161/1998, and by *Institut des Hautes Études Scientifiques* (Bures-sur-Yvette, France).